\documentclass[ngerman,english]{bvm} 

\bvmdef\articlenumber{3046}

\bvmdef\type{P}

\date{}

\title{Deep Learning compatible Differentiable X-ray Projections for Inverse Rendering}

\titlerunning{Differentiable X-ray Projections for Inverse Rendering}

\author{Karthik~Shetty$^{1}$, Annette~Birkhold$^2$, Norbert~Strobel$^{3}$, Bernhard~Egger$^{4}$, Srikrishna~Jaganathan$^{1}$, Markus~Kowarschik$^{2}$, Andreas~Maier$^1$}

\authorrunning{Shetty et al.}

\institute{%
$^1$Pattern Recognition Lab, FAU Erlangen-N\"urnberg\\
$^2$Siemens Healthcare GmbH, Forchheim\\
$^3$Fakult\"at Elektrotechnik, HS f\"ur angewandte Wissenschaften W\"urzburg-Schweinfurt\\
$^4$MIT - BCS, CSAIL \& CBMM, USA}

\email{karthik.shetty@fau.de}

\begin{document}

%
\selectlanguage{english}

\maketitle
\begin{abstract}
Many minimally invasive interventional procedures still rely on 2D fluoroscopic imaging. Generating a patient-specific 3D model from these X-ray projection data would allow to improve the procedural workflow, e.g. by providing assistance functions such as automatic positioning. To accomplish this, two things are required. First, a statistical human shape model of the human anatomy and second, a differentiable X-ray renderer. In this work, we propose a differentiable renderer by deriving the distance travelled by a ray inside mesh structures to generate a distance map. To demonstrate its functioning, we use it for simulating X-ray images from human shape models. Then we show its application by solving the inverse problem, namely reconstructing 3D models from real 2D fluoroscopy images of the pelvis, which is an ideal anatomical structure for patient registration. This is accomplished by an iterative optimization strategy using gradient descent. With the majority of the pelvis being in the fluoroscopic field of view, we achieve a mean Hausdorff distance of 30\ts mm between the reconstructed model and the ground truth segmentation.

\end{abstract}

\section{Introduction}

Over the last decades, the amount of X-ray guided interventional procedures has increased steadily, raising the awareness for optimized procedural workflows and radiation-dose induced adverse effects. Assistance systems based on a 3D digital twin of the patient have the potential to speed up procedures and minimize the required radiation. However, precisely representing the individual human anatomy based on the commonly acquired 2D projection images is an ill-posed problem. Ehlke et al. showed that an accurate and fast 3D reconstruction of the human pelvis from 2D projection images is possible by learning its shape space in the form of tetrahedral mesh structures from a large set of pelvis scans including bone density information~\cite{3046-07}. In theory, this approach has the potential to generate digital twins of patients from live fluoroscopic imaging data and may allow to develop advanced assistance applications in the interventional environment.  

The process of generating a 3D model from 2D projection images is known as inverse-rendering and has received considerable attention in computer vision~\cite{3046-02,3046-05}. However, achieving this usually means solving an ill-posed problem, and rendering pipelines are not necessarily differentiable. This is usually overcome by softening the non-differentiabilities by soft-rasterization~\cite{3046-02}. Usually, a 3D scene is  represented by a polygon mesh, which can be expressed by a set of vertices and faces. Such a representation is usually chosen for statistical shape and pose modeling. Skinned Multi-Person Linear Model (SMPL) is an example of such a parameterized model, which is deformable and accurately represents a large variation of human shapes~\cite{3046-03}. Employing such a model enables more precise and faster 3D reconstruction from a 2D projection image space due to the incorporation of prior knowledge~\cite{3046-05}. 
To carry out the 3D reconstruction from X-ray projections, we need to calculate artificial X-ray projections from  a 3D model, e.g., a Statistical Shape Model (SSM). Generation of transmission images from mesh structures has been previously proposed \cite{3046-07,3046-01}. For a viable reconstruction, we seek to optimize the shape $\vec\beta$ and pose $\vec\theta$ parameters by minimizing a similarity function between the simulated image and the acquired image. To this end, we propose a framework to combine existing X-ray rasterization approaches with a differentiable rasterizer, to create a setup that can be iteratively optimized by gradient descent techniques. The proposed renderer enables future end-to-end frameworks.

\section{Methods}
\subsection{Differentiable Rasterizer}\label{3046-raster}

\begin{figure}[htb]
\centering
\setlength{\figbreite}{0.5\linewidth}
\includegraphics[width=\figbreite,height=\figbreite,keepaspectratio]{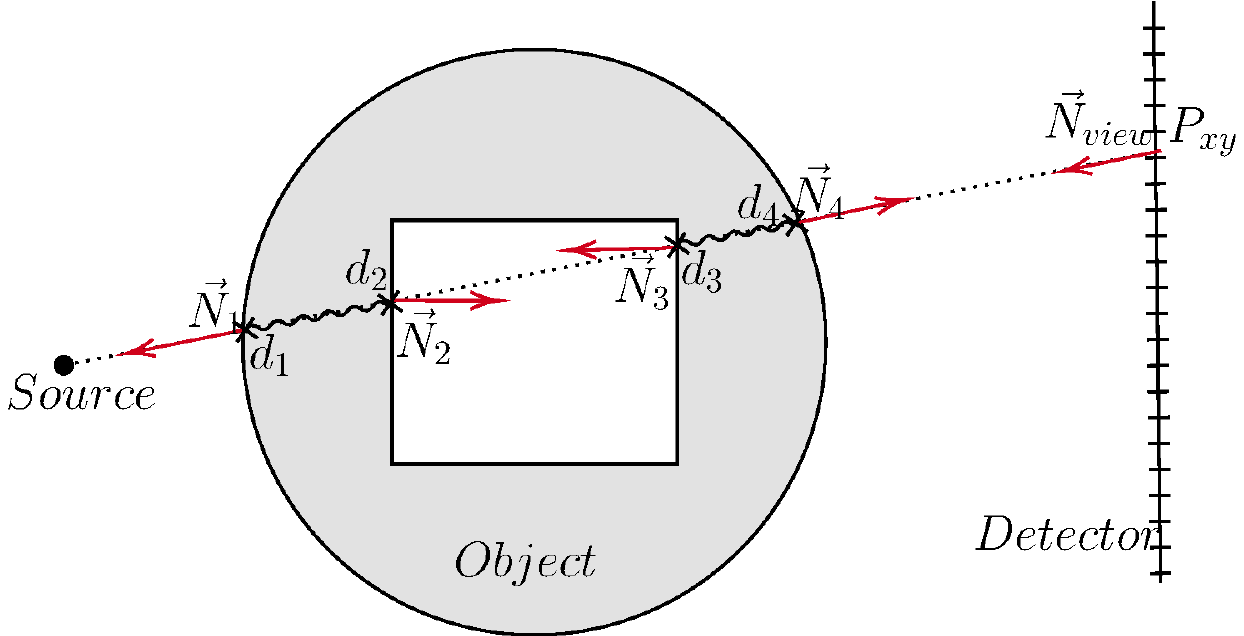}
\caption{Visualization for the determination of the ray distance for a simplified structure. Here, total distance covered by the ray is $L_{xy}\textrm{=}d_1\textrm{-}d_2\textrm{+}d_3\textrm{-}d_4$.} 
\label{3046-fig-001}
\end{figure}

A watertight manifold surface can be represented as a triangular mesh by a set of faces $\boldsymbol{f} \in \mathbb{N}^{N_f \times 3}$ and vertices $\boldsymbol{v} \in \mathbb{R}^{N_v \times 3}$, where the the object has $N_f$ faces and $N_v$ vertices. Given a projection matrix, it is possible to transform the mesh from world coordinates to detector coordinates while maintaining the actual depth of the vertex. In traditional rasterization of 3D scenes, each pixel $P_{xy}$ on the screen is affected by only the face $\boldsymbol{f}_j$ nearest to the ray from the camera. This is usually achieved by storing the depth information for all the faces influenced by a given pixel in the form of \emph{z-buffer}. However, in the case of transmission imaging we need to know the distance traveled by a ray through an object in the form of \emph{l-buffer}~\cite{3046-01}. For each face $\boldsymbol{f}_j$, the normals $\overrightarrow{N}_{\boldsymbol{f}_j}$ in the projective space are computed. 
The sign of the dot product $D_{\boldsymbol{f}_jR_{xy}}
 = sign\left(\overrightarrow{N}_{\boldsymbol{f}_j} \cdot  \overrightarrow{N}_{view} \right)$ between the face normal $\overrightarrow{N}_{\boldsymbol{f}_j}$ and detector plane normal $\overrightarrow{N}_{view} = \left(0, 0, 1\right)^T$ 
determines if the ray $R_{xy}$ is either entering or exiting the object for a given face. Hence, the total path length $L_{xy}$ traveled by a ray for a given object $p$ from the source to the detector pixel position $P_{xy}$ is visualized in Fig.~\ref{3046-fig-001} and can be formulated as   
\begin{equation}
\label{3046-eq-02}
L_{xy} = \sum_{j=1}^K D_{\boldsymbol{f}_jR_{xy}} Z_{xy\boldsymbol{f}_j}
\end{equation}

Here $Z_{xy\boldsymbol{f}_j}$ represents the distance between the face $\boldsymbol{f}_j$ and the detector pixel position $P_{xy}$, while $j$ and $K$ represents the current valid intersection and the fixed number of mesh overlaps stored in the z-buffer, respectively. As a result, there is no substantial requirement for the elements of the z-buffer to be stored in a sorted manner. The only constraint is the number of faces $K$ influencing a given pixel.
To determine the gradients for the backward pass we follow the same procedure as implemented by Ravi et al \cite{3046-02}, with an exception for the gradients for the z-buffer:
\begin{equation}
\label{3046-eq-03}
\frac{\partial C}{\partial z_{xyj}} = \frac{\partial C}{\partial L_{xy}}D_{\boldsymbol{f}_jR_{xy}}
\end{equation}
Here, $z_{xyj}$ represents the z-buffer distance from the pixel $P_{xy}$ corresponding to face $\boldsymbol{f}_j$ with $C$ as some cost function, e.g., the normalized gradient correlation (NGC) explained below. 

It is generally assumed that the number of incoming rays for a given object is the same as the outgoing rays. However, artifacts occur when the face normal is perpendicular to the detector plane or in the presence of a non-manifold surface. The occurrence of such a situation can be determined when the summation over the pixel sign function $\sum_j^K R_{xy\boldsymbol{f}_j}$ is non-zero. To overcome this, we set the image pixel value to a neighbouring pixel with zero sum and set the gradient for that particular pixel to zero.

\subsection{X-ray rendering}\label{3046-render}
\begin{figure}[htb]
	\setlength{\figbreite}{0.28\textwidth}
	\centering
	\subfigure{\includegraphics[width=\figbreite,height=\figbreite,keepaspectratio]{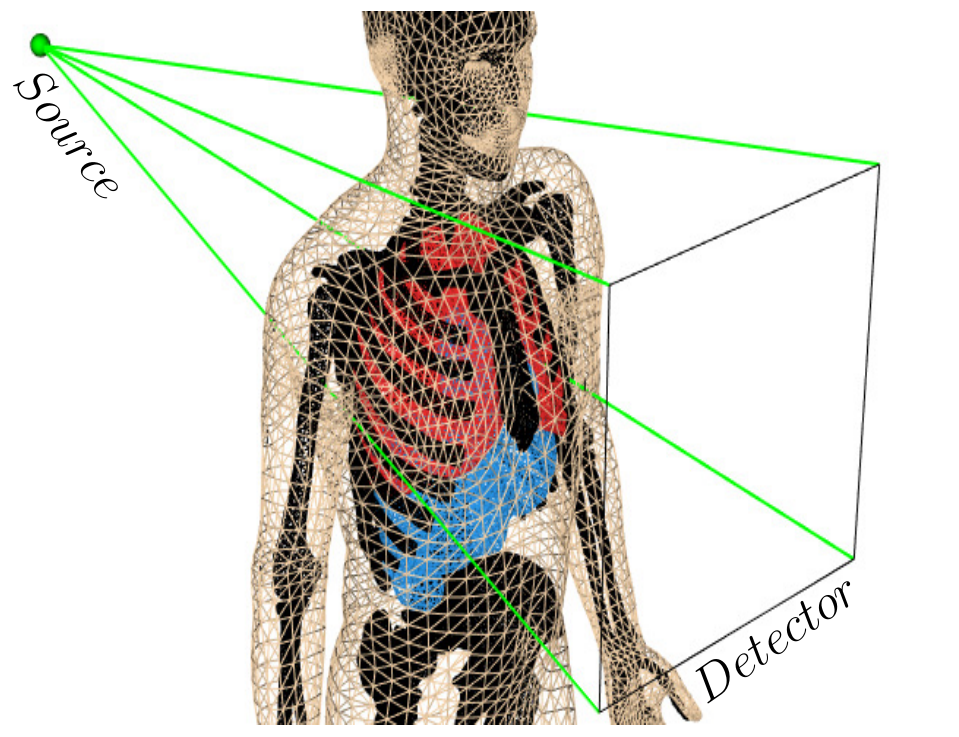}}
	\setlength{\figbreite}{0.65\textwidth}
	\subfigure{\includegraphics[width=\figbreite,height=\figbreite,keepaspectratio]{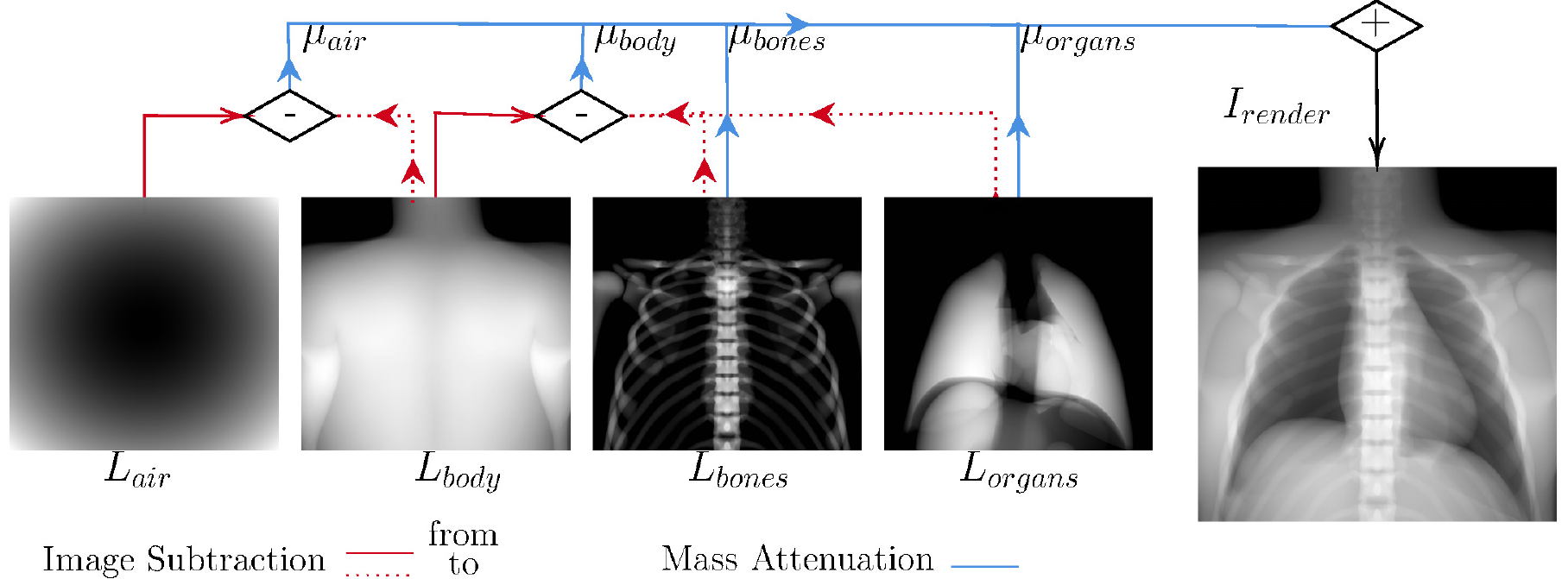}}
	\caption{Left: Projection performed around the thorax section from a mesh representation of a patient model. Right: Pipeline to generate a transmission image from the given mesh structures and projection matrix.}
	\label{3046-fig-002}
\end{figure}

The pipeline for generating the X-ray projection image for a human patient model is depicted in Fig.~\ref{3046-fig-002}. Distance projection maps $L_{p}$ are generated individually for air ($L_{air}$), body ($L_{body}$), bones ($L_{bones}$) and organs ($L_{organs}$). $L_{air}$ represents the Euclidean distance from a detector pixel $P_{xy}$ in world coordinates to the X-ray source. $L_{body}$, $L_{bones}$ and $L_{organs}$ are generated as described in Sec.~\ref{3046-raster} from individual meshes respectively. For ease of explanation, we describe $L_{organs}$ as a single organ distance map, however it is a subset of lungs, heart, liver, kidney and spleen. As $L_{air}$ is inclusive of $L_{body}$, we define the actual air distance map as $L_{air} - L_{body}$. For the body we proceed similarly as shown in Fig.~\ref{3046-fig-002}.

The X-ray projection is generated using a primary signal $I_{xy}$ for a pixel $P_{xy}$ described using the Beer-Lambert law as shown in Eq.~\ref{3046-eq-04} with a photon energy $E$, X-ray energy $I_{o}(E)$, and linear attenuation coefficients $\mu(p, E)$ for objects $p$

\begin{equation}
\label{3046-eq-04}
I_{xy}= \sum_E I_{o}(E) e^{\sum_p \left( \mu(p, E)L_{p_{xy}} \right)}
\end{equation}
Making use of the aforementioned distance maps $L_p$, the simulation can be extended for polychromatic X-ray beam spectra.

\subsection{Application: Inverse Rendering}\label{3046-application}
We build a statistical parametric model $M(\vec{\beta}, \vec{\theta}; \mathbf{\Phi})$, similar to SMPL~\cite{3046-03} extended for internal anatomy using principal component analysis to characterize the human anatomy shape space from a large set of segmented whole-body CT scans. This makes it possible to generate realistic X-ray projections with the entire pipeline being differentiable in nature. Here, 
$\mathbf{{\Phi}}$ represents the learned parameters of the model. To reconstruct a 3D model from a projection image we fix the camera projection matrix of the simulation system as defined by the target along with an initial mean shape $\vec{\beta}_m$ and rest pose $\vec{\theta}_r$. A gradient descent based optimization with backpropagation is applied to the simulation model. NGC is used as a similarity measure \cite{3046-04}, which is defined by the normalized cross-correlation of the image gradient between two images. The model is optimized until convergence to obtain the optimal pose $\vec{\theta}$ and shape $\vec{\beta}$ parameters.

For evaluation, we used the dataset provided by Grupp et al. made up of the hip anatomy from 6 patients \cite{3046-08}. For each patient, a 3D segmentation of the pelvis along with 14 landmarks and a total of 366 fluoroscopy images in varying orientations with their respective projection matrices are available. 

\section{Evaluation and Results}
\begin{figure}[htb]
	\setlength{\figbreite}{0.155\textwidth}
	\centering
	\subfigure[]{\includegraphics[width=\figbreite,height=0.3268\textwidth]{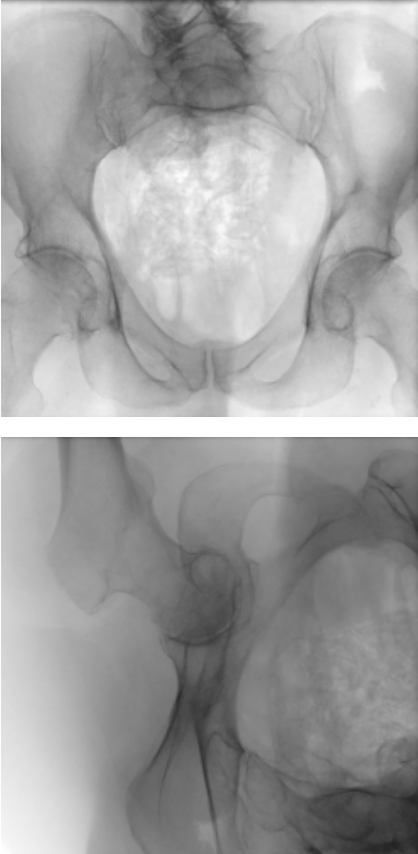}}
	\subfigure[]{\includegraphics[width=\figbreite,height=0.3268\textwidth]{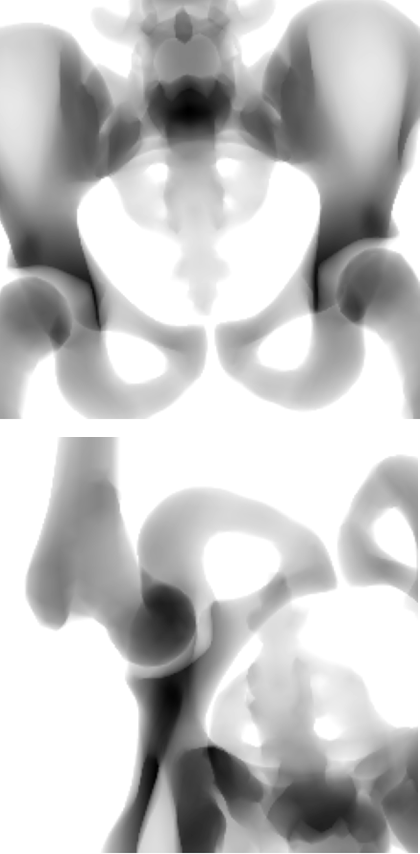}}
	\subfigure[]{\includegraphics[width=\figbreite,height=0.3268\textwidth]{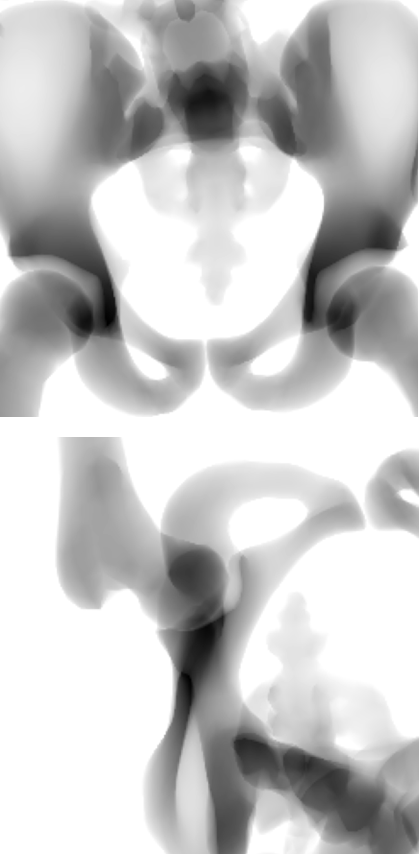}}
	\subfigure[]{\includegraphics[width=\figbreite,height=0.3268\textwidth]{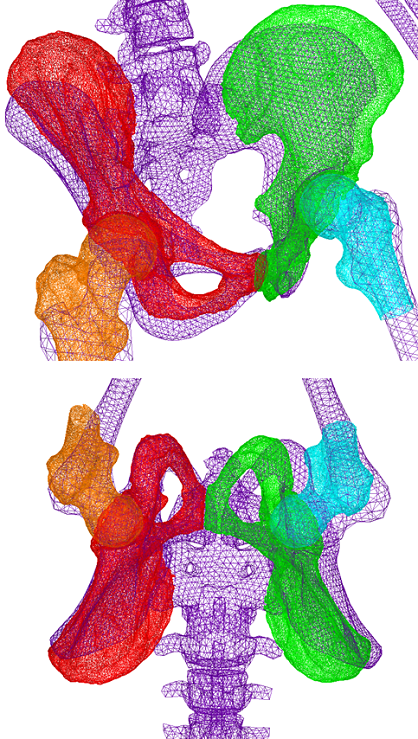}}
	\subfigure[]{\includegraphics[width=\figbreite,height=0.3268\textwidth]{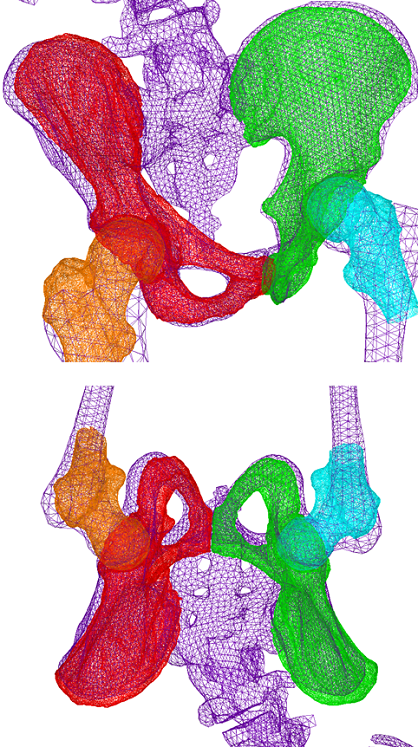}}
	\subfigure[]{\includegraphics[width=0.185\textwidth,height=0.3268\textwidth]{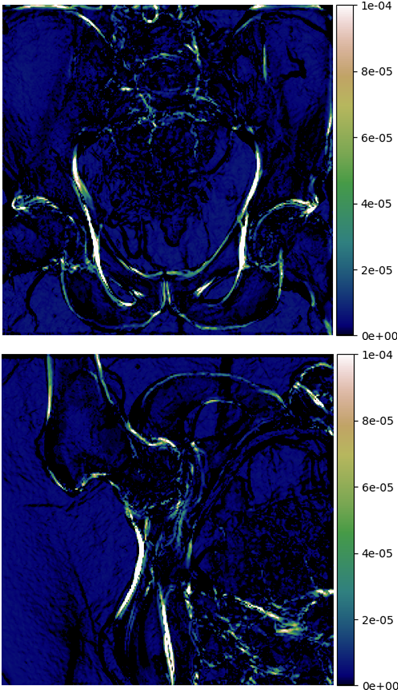}}
	\caption{Samples for 3D reconstruction from 2D projection images. (a) Target image, (b) Projection image with a random translation and rotation, (c) Projection image after registration, (d) Initial 3D overlay of template mesh, (e) 3D overlay of the meshes after registration, (f) NGC map after registration.}
	\label{3046-fig-003}
\end{figure}

The SSM is registered to each patient from the dataset. This resulted in a base score with an average Hausdorff distance of 17.79\ts mm and a mean landmark error of 11.81\ts mm. The rotation and translation parameters obtained are assumed as the ground truth irrespective of the shape parameters. The 2D-3D reconstruction is tested by applying a random translation in the range of [-10,10]\ts mm and a rotation in the range of [-5,5]$^{\circ}$ on all three axes for the SSM. The randomization is small due to the nature of the loss function being sensitive to changes only around a small region, such precision of initialization could be either provided manually or achieved with a network learning those initialization parameters. Fig.~\ref{3046-fig-003} shows sample outputs of our proposed method along with the quantitative results in Table~\ref{3046-tab-001}. For evaluation, we consider only the images where the region of pelvis visible is greater than $50\%$. 

\begin{table}[]

\caption{Overview of the reconstruction and landmark error. All distances are measured in mm. The lower bound error is dependent on the shape model, with a mean Hausdorff distance of 17.79\ts mm and a mean landmark error of 11.81\ts mm.}
\begin{tabular*}{\textwidth}{c@{\extracolsep\fill}cccc}
\hline

\begin{tabular}[c]{@{}c@{}}Initial \\ Hausdorff Distance\end{tabular} & \begin{tabular}[c]{@{}c@{}}Initial \\ Landmark Error\end{tabular} & \begin{tabular}[c]{@{}c@{}}Reconstructed\\  Hausdorff Distance\end{tabular} & \begin{tabular}[c]{@{}c@{}}Reconstructed \\ Landmark Error\end{tabular} \\
 \hline
$46.59\pm4.64$                 & $30.00\pm8.30$               &$29.11\pm4.56$                      & $22.06\pm6.13$                   \\
 \end{tabular*}
\label{3046-tab-001}
\end{table}

\section{Discussion and Conclusion}
We presented an approach for differentiable rendering and reconstructing a patient-specific 3D mesh model from a single 2D X-ray projection image. We demonstrated that using a deformable human model parameterized by shape and pose, we were able to approximately register patients to an X-ray projection. 
We found that given a visibility above 50\% and an initialization of a generic model shape positioned close to the actual image, the 3D  reconstruction error is significantly reduced with our optimization approach. The remaining error may be due to the simultaneous unconstrained optimization of shape and pose. This is because changes in shape and translation across depth, i.e., in z-direction, results in a large range of solutions causing the simulated image to appear similar to the target image. For now, we ignore the case when less than $50\%$ of the pelvis is visible due to the intrinsic nature of the loss function being sensitive to the initial position. In general, the error could be reduced with an additional prior to determine a good initial pose~\cite{3046-10}.

As the registration is independent of the patient pose we can overcome possible pose mismatches resulting from employing preoperative CT scan for registration. Still, the pipeline has limitations. First, the X-ray simulation is an approximation of a real X-ray image, as it does not take into account any scattering mechanisms and other effects causing noise. This can be possibly improved by employing a deep learning model for predicting patient-specific scatter~\cite{3046-09}. Second, the renderer assumes a constant density for any given material, resulting in a pseudo-realistic rendering. One way to overcome this problem would be to use a tetrahedral mesh with learned bone density distributions.
Third, due to ambiguous depth information in fluoroscopy images, accurate registration becomes a challenging  task.
This error could be minimized with multi-view images or by using the table position as a constraint. To summarize, we proposed a differentiable renderer by deriving the distance travelled by a ray inside mesh structures. This was further extended to generate X-ray images from human models. Finally, we showed that it could be successfully used to reconstruct 3D models from 2D  fluoroscopy images.

\paragraph{Disclaimer:}
The concepts and information presented in this paper are based
on research and are not commercially available. 

\bibliographystyle{bvm}

\bibliography{3046}

\begin{thebibliography}{1}

\bibitem{3046-07}
Ehlke M, Ramm H, Lamecker H, et~al.
\newblock Fast generation of virtual X-ray images for reconstruction of 3D
  anatomy.
\newblock IEEE Trans Vis Comput Graph. 2013; p. 2673--2682.

\bibitem{3046-02}
Ravi N, Reizenstein J, Novotny D, et~al.
\newblock Accelerating 3d deep learning with pytorch3d.
\newblock arXiv preprint arXiv:200708501. 2020;.

\bibitem{3046-05}
Liu S, Li T, Chen W, et~al.
\newblock Soft rasterizer: A differentiable renderer for image-based 3d
  reasoning.
\newblock In: Proc IEEE Int Conf Comput Vis; 2019.  p. 7708--7717.

\bibitem{3046-03}
Loper M, Mahmood N, Romero J, et~al.
\newblock {SMPL}: A Skinned Multi-Person Linear Model.
\newblock ACM Trans Graph (Proc SIGGRAPH Asia). 2015 Oct;34(6):248:1--248:16.

\bibitem{3046-01}
Vidal FP, Garnier M, Freud N, et~al.
\newblock {Simulation of X-ray Attenuation on the GPU}.
\newblock In: Tang W, Collomosse J, editors. Theory and Practice of Computer
  Graphics. The Eurographics Association; 2009. .

\bibitem{3046-04}
Penney GP, Weese J, Little JA, et~al.
\newblock A comparison of similarity measures for use in 2-D-3-D medical image
  registration.
\newblock IEEE Trans Med Imaging. 1998;17(4):586--595.

\bibitem{3046-08}
Grupp RB, Unberath M, Gao C, et~al.
\newblock Automatic annotation of hip anatomy in fluoroscopy for robust and
  efficient {2D}/{3D} registration.
\newblock Int J Comput Assist Radiol Surg. 2020 May;15(5):759--769.

\bibitem{3046-10}
Bier B, Unberath M, Zaech JN, et~al.
\newblock X-ray-transform invariant anatomical landmark detection for pelvic
  trauma surgery.
\newblock In: Med Image Comput Comput Assist Interv. Springer; 2018.  p.
  55--63.

\bibitem{3046-09}
Roser P, Zhong X, Birkhold A, et~al.
\newblock {Physics}-driven learning of x-ray skin dose distribution in
  interventional procedures.
\newblock Med Phys. 2019;46:4654--4665.

\end{thebibliography}
\marginpar{\color{white}E\articlenumber} 
\end{document}